\documentclass[amsmath,amssymb,amsfonts,twocolumn]{revtex4}
\usepackage{graphicx}
\usepackage{bbm}
\usepackage{float}
\usepackage{latexsym}
\usepackage{braket,bigints}
\usepackage{amsmath,stmaryrd,mathtools}
\def \beq {\begin{equation}}
\def \eeq {\end{equation}}

\begin{document}

\title{Spatially-selective and quantum-statistics-limited light stimulus for retina biometrics and pupillometry}
\author{A. Margaritakis\thanks{First two authors contributed equally} , G. Anyfantaki, K. Mouloudakis,  A. Gratsea and I. K. Kominis}
\affiliation{Department of Physics, University of Crete, 71003 Heraklion, Greece\\ Institute for Theoretical and Computational Physics, University of Crete, 70013 Heraklion, Greece}
\begin{abstract}
Quantum vision is currently emerging as an interdisciplinary field synthesizing the physiology of human vision with modern quantum optics.
We recently proposed a biometric scheme based on the human visual system's ability to perform photon counting. We here present a light stimulus source that can provide for the requirements of this biometric scheme, namely a laser light beam having a cross section consisting of discrete pixels, allowing for an arbitrary pattern of illuminated pixels, each having a photon number per unit time obeying a Poisson distribution around the mean. Both the illumination pattern and the photon number per pixel per unit time are computer controlled, offering simple and unsupervised scanning of these parameters. Moreover, infrared light exactly superimposed on the stimulus pattern can be used for acquiring exact information on the illumination geometry on the pupil. The same light stimulus source can be used to advance pupillometry to a higher level of control and precision, as current pupillometers lack spatial selectivity in illumination. We present pupillometry measurements demonstrating the potential of this source to offer a wealth of information on vision and brain function.
\end{abstract}
\maketitle
\section{Introduction}
Quantum vision, the synthesis of quantum optics with the physiology of human vision is a very promising new research direction falling into the broader study of quantum effects in biological systems \cite{Plenio_review,Kominis_review,Bowen}. Exciting developements in this synthesis have been studies of the response \cite{Baylor,Rieke1,Rieke2,Rieke3,Rieke4,Reingruber,Pugh} of single (mammalian or amphibian) rod cells to single photons produced by modern single-photon sources \cite{Kr1,Kr2,Kr3}. Alongside several proposals to test quantum mechanics with the human eye \cite{Gisin1,Gisin2,Hornberger,Pizzi,Vivoli,Dodel}, recent experiments with humans \cite{Vaziri,Holmes} also utilized a single photon source to test the single-photon sensitivity and integration time of the human visual system \cite{Nelson}, both being fundamental questions already under investigation with classical light sources since the 1930s and onwards \cite{Vavilov,Hecht,deVries,Rose,Sakitt}. 

From an applied physics aspect, it was recently proposed \cite{Kominis} to use measurements at the human visual threshold to indentify human subjects, a method termed quantum biometrics, as it relies on the statistics of photodetection of few-photon coherent light beams. In particular, the methodology of this proposal is based on measuring the optical loss light suffers along its path to the retina along several distinct paths illuminating several distinct retinal spots. These losses can be measured by illuminating the eye with a coherent light pulse having a known and controllable photon number, and interrogating the subject on the perception of the light pulse. The optical losses are quantified by a parameter $\alpha$, where $0\leq\alpha\leq 1$, in particular by a whole retinal map of $\alpha$ parameters. By changing the incident photon number and registering the human responses, one can scan the probability of seeing curve from 0 to 1 and infer the individual's $\alpha$-map. Subsequent measurement protocols can swiftly identify individuals with quantum-limited security and an unprecedented performance in terms of false-positive probability \cite{Kominis}.

In this work we report on the development of the light stimulus source that allows the realization of the aforementioned biometric protocol. In particular, the source can provide for any pattern of laser light pulses consisting of up to 25 pixels on a $5\times 5$ grid, each illuminated pixel having a Poisson distributed photon number with mean ranging from 10 to 200 photons. Importantly, both the illumination pattern and the photon number are computer controlled, as required for using the source in automatically applying measurement protocols with the least amount of time from the side of the device's operator, and thus the least amount of effort from the side of the interrogated subject.

Interestingly, the same source, albeit with more intense pulses, can be used to advance pupillometry \cite{nowak,chilcott,mathot} towards unraveling yet inaccesibe information, and among several possibilities, open a new window into the human brain. Indeed, pupillometry studies the dynamics of pupil diameter changes under illumination and under various physiological conditions, and has been extensively used as a diagnostic tool in medicine. The spatially-selective light stimulus source reported herein will thus allow to resolve finer detail in a number of pupillometry studies on opthalmology, neurology, pharmacology and brain function. 

The outline of the paper is the following. In Section 2 we present the detailed physical requirements of the light stimulus source to be used in the aforementioned vision studies. In Section 3 we present the experimental apparatus of the light stimulus source and the measurements characterizing its performance. In Section 4 we demonstrate the use of the light stimulus source in spatially selective pupillometry, and conclude with an outlook in Section 5.
\section{Physical requirements for a light stimulus used in biometrics and pupillometry}
In this section we first reiterate the idea behind the quantum biometrics methodology introduced in \cite{Kominis}, and then define the requirements for the light stimulus source realizing this methodology. Subsequently we briefly introduce pupillometry and make the case that the same light stimulus source can be used to advance the state-of-the-art of this medical diagnostic technique.
\subsection{Quantum biometrics with photon counting by the retina}
\begin{figure}
\begin{center}
\resizebox{0.45\textwidth}{!}{
 \includegraphics{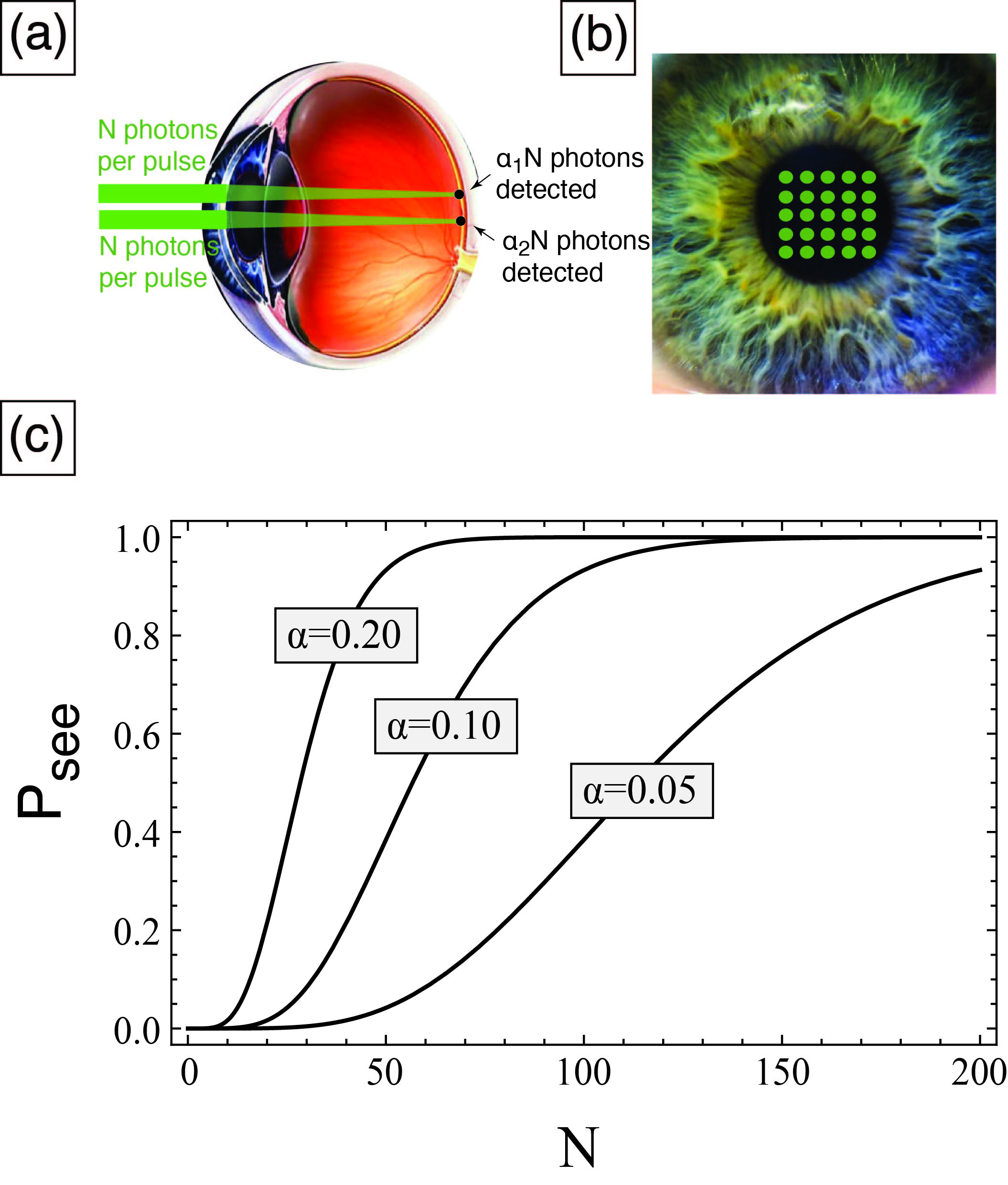}
}
\end{center}
\caption{(a) Biometric methodology introduced in \cite{Kominis}. Light incident on the eye on different spots travels towards the retina along different paths suffering different optical losses (which include the probability of no detection at the particular retinal spot being illuminated). These losses are quantified by a parameter $\alpha$, which depends on the particular path taken by the light beam towards the retina. The two black spots in the figure define the points where the two parallel beams characterized by the parameters $\alpha_1$ and $\alpha_2$ cross the retina. The figure is a schematic illustration rather than a precise geometrical optics description, in the sense that for a normal eye the two black spots coincide, whereas for a myopic eye the focus precedes the retina and the two black spots are indeed in different positions, as shown in the figure. (b) Regardless of the type of eye considered, measuring these optical losses on a number of pixels incident on the pupil should lead to an $\alpha$-map characteristic of the individual. (c) The measurement of each $\alpha$ value of the laser beam grid is performed pixel by pixel by changing the incident photon number and interrogating the human subject on the perception or not of the light. Shown here are indicative probability-of-seeing curves as a function of the incident photon number per pulse, $N$, for three different values of $\alpha$.  $P_{\rm see}=\sum_{n=K}^{\infty}p(n;\alpha N)$, with $p(n;\overline{n})=e^{-\overline{n}}\overline{n}^n/n!$, i.e. a sum over a Poisson-distributed photon number for all possible numbers of detected photons exceeding the visual preception threshold of $K=6$ (see \cite{Kominis,Bialek}).}
\label{biom}
\end{figure}

The workings of the method introduced in \cite{Kominis} are summarized in Fig. \ref{biom}. If we illuminate the eye with a pulse of a coherent light beam (duration $\tau$) having $N$ photons per pulse on average, then due to the optical losses light suffers along its path towards the retina, only $\alpha N$ photons will be detected, where $0\leq\alpha\leq 1$ is the optical loss parameter. Typical values for $\alpha$ reach up to 0.2.  Thus, as shown in Fig. \ref{biom}(a), different light paths will suffer different losses, leading to a different number of detected photons. Losses include the probability of photon detection at the illuminated retinal spot. Parenthetically we note that the geometry of Fig. \ref{biom}(a) is not to be taken literally, since the illumination geometry {\it on the retina} of what starts as a number of parallel light paths incident on different spots {\it on the pupil} will depend on whether the eye is normal, nearsighted or farsighted, plus on additional optics that can be introduced before the eye.

In any case, if we can shine light on several different spots on the pupil, like in Fig. \ref{biom}(b), we can form an $\alpha$-map characteristic for each individual. In the schematic example of Fig. \ref{biom}(b) we have a $5\times 5$ grid of illuminated pixels on the pupil, thus forming an $\alpha$-map having 25 entries. To measure each $\alpha$ we have to scan the probability of seeing curve, shown in Fig. \ref{biom}(c). 

This probability can be calculated by considering the Poissonian photon statistics in the light pulses having average photon number $N$, as well as the detection threshold of the visual system. By varying $N$ and interrogating the human subject on the perception or not of the light pulses, and doing so for each of the 25 illumination pixels, one can infer the $\alpha$-map for the particular example of Fig. \ref{biom}b.
This forms the first and time-consuming step of registering the biometric characteristics of a particular individual. The identification process is much faster and rests on classifying the $\alpha$ values in high- and low-$\alpha$. When the subject wishes to be identified, one can for example simultaneously illuminate two spots having very low $\alpha$ together with three spots having very high $\alpha$, and asking the individual to respond on how many light spots he/she perceived, given a range of possible answers, e.g. from 0 to 10. If the individual taking the test is the one he/she claims to be, the answer would be three with a high probability. If it is an impostor impersonating somebody else, the impostor has as an only option to answer randomly, since all pixels are illuminated with the same number of photons and no information can thus be inferred on the $\alpha$-map of the person being impersonated. In this particular example, the probability for the impostor to get the correct answer is 0.1. By repeating a few such tests one can reject impostors with any desired false positive probability. 

Thus, to realize these measurements we need to be able to shine laser light beams in an arbitrary pattern illuminating the pupil, i.e. any number of pixels inside a grid of $n\times n$ pixels, which we henceforth call illumination pixels. The average photon number per pulse per pixel should be the same for all illuminated pixels, and be controllable from a small value where the probability of seeing, $P_{\rm see}$, is small to a value where $P_{\rm see}\approx 1$. Referring to Fig. \ref{biom}(c), a reasonable range for the photon number is from 20 to 200. Furthermore, to actually realize the most precise estimate of $\alpha$ the photon number per pulse per pixel should have the smallest possible variance, which for coherent light is equal to the average photon number, i.e. the photon number follows the Poisson distribution. Finally, the photon number and illumination pattern should be computer controlled in order to maximize the speed of the data acquisition and the comfort of the interrogated subject. 

Before proceeding with the presentation of how we realized the aforementioned light stimulus source, we describe how the same source can advance the precision of a broadly used medical diagnostic technique, pupillometry. 
\subsection{Pupillometry}
\begin{figure}
\resizebox{0.45\textwidth}{!}{
 \includegraphics{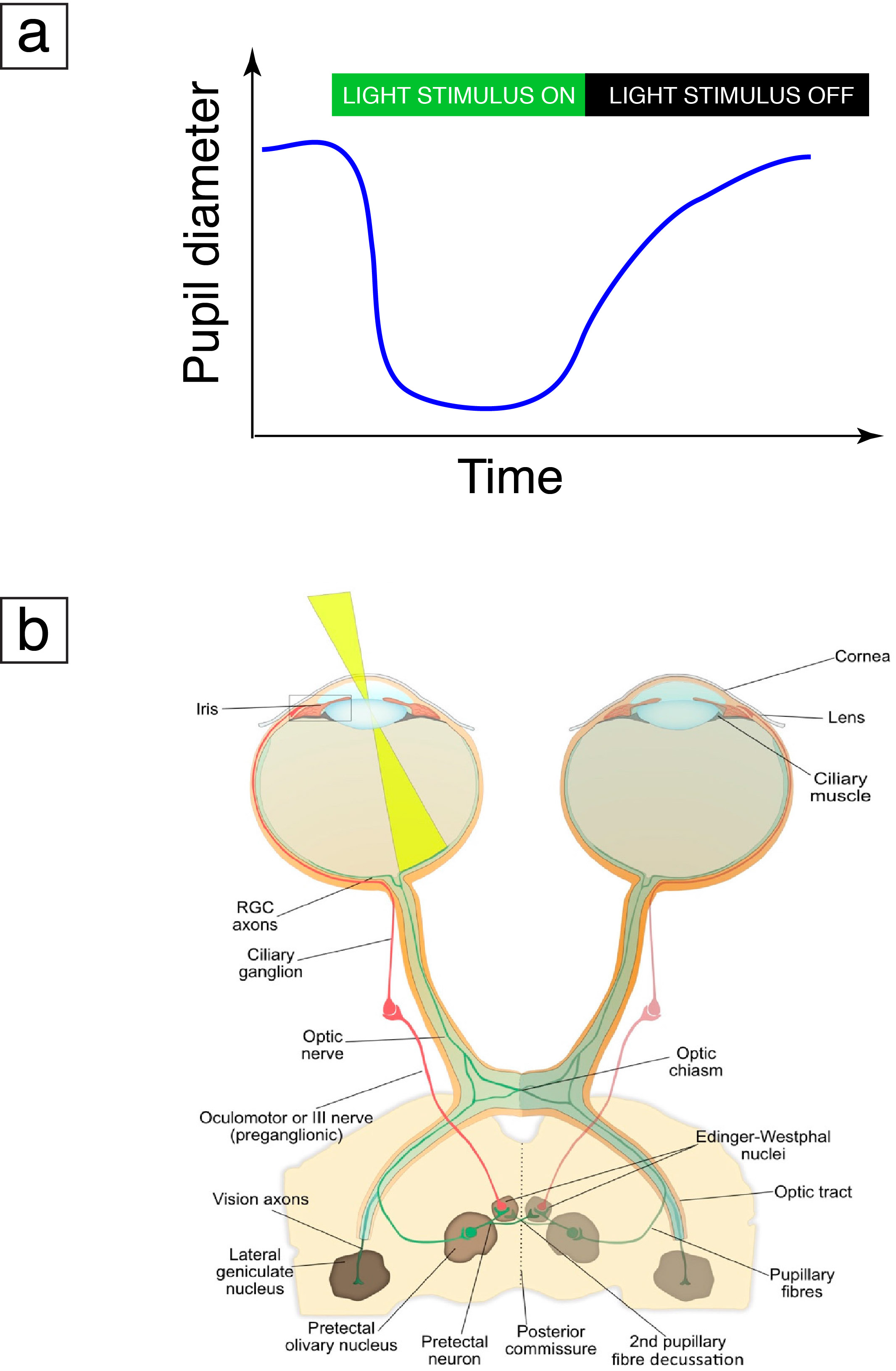}
}
\caption{(a) Pupillometry is the measurement of the pupil diameter under various illumination and physiological conditions. (b) Besides its medical diagnostic value, pupillometry can probe several neural circuits (figure adapted from \cite{chilcott}). }
\label{per}
\end{figure}
Pupillometry is about measuring the dynamical changes of pupil diameter upon various illumination and physiological conditions. A typical pupil light reflex is shown in Fig. \ref{per}(a). As shown in Fig. \ref{per}(b), the dynamics of the pupil are 
regulated by elaborate neural circuits of the sympathetic and parasympathetic nervous system. So far, the light stimulus used in pupillometry is a classical light source (lamp or led) illuminating a significant part of the pupil. Precisely measuring the temporal variation of pupil size under spatially selective illumination conditions can thus open a new window to studies of brain function as well as understanding and diagnozing several neurological diseases \cite{stark,AJO1964,longtin,howarth1,howarth2,wilhelm1,wilhelm2,ricarte}.
\subsection{Light stimulus source of this work}
The light stimulus source we develop here can realize the requirements of the biometric methodology previously described. Additionally, used with a higher intensity, it can provide for pupillometry studies having spatial selectivity. 
In the following we will describe the development of a laser light beam, the cross section of which consists of discrete pixels that can be illuminated or non-illuminated, here in the geometry of a $5\times 5$ grid. 

The photon number per unit time per illuminated pixel, as well as the specific pattern of illuminated pixels (from one to all 25) can be computer-controlled. Due to the coherence properties of laser light, the photon number is precisely known, as it is Poisson distributed around its mean value. Additionally, an important aspect of the source is that it involves infrared light exactly superimposed on the green stimulus light, so that an infrared camera can monitor, through reflection of the infrared light, the exact geometry of illumination on the pupil. 

Other works on studying vision with spatial selectivity involve single-photoreceptor cell illumination \cite{Harmening,Sabesan,Marcos} using a broadband supercontinuum laser source containing both stimulus light and infrared pointing light. The spatial resolution of this methodology is at the single-cell level, however it is achieved at the expense of a significantly more elaborate apparatus of a more "invasive" nature. Regarding pupillometry, to our knowledge there has been so far a single yet very thorough study \cite{Wyatt} having only a gross spatial selectivity, i,e. using classical light to target nasal versus temporal retina.  
\section{Experimental apparatus}
\begin{figure*}
\begin{center}
\resizebox{0.95\textwidth}{!}{
 \includegraphics{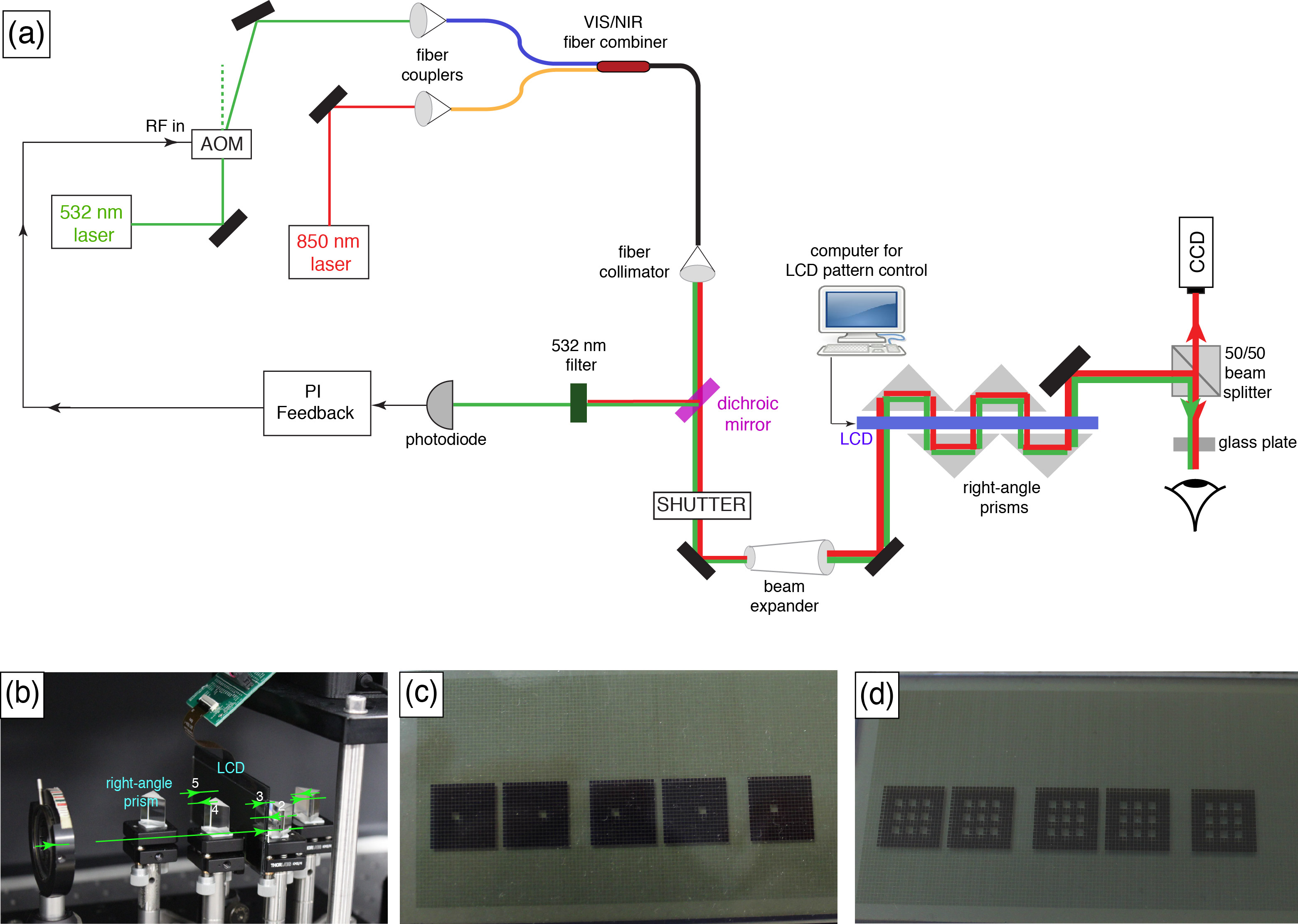}
}
\end{center}
\caption{(a) The experimental setup for delivery to the human eye of a desired pattern of pixels having mean photon number between 20 and 200 photons per integrating pulse of 100 ms (see text for a detailed description). The 532 nm laser is used for the stimulus light, while the 850 nm laser for monitoring the geometry of stimulation with the camera, using the infrared light reflection by a glass plate preceding the eye. Both wavelenghts are combined into a fiber and thus propagate towards the eye in a common mode. The intensity of the 532 nm light exiting the fiber combiner is actively stabilized using the rf power of the AOM. The illumination pattern is formed by a multi-pass configuration through an LCD. (b) Picture of LCD with the right-angle prisms used to create the indicated 5 passes through the same pattern of activated/inactivated LCD dots. (c) An example of an LCD dot pattern creating just one 532 illumination pixel among the 25 possible pixels in the $5\times 5$ grid. (d) Another example creating a $3\times 3$ illumination grid.}
\label{setup}
\end{figure*}
The experimental apparatus is shown in Fig. \ref{setup}. The stimulus and pointing light is provided by a 50 mW cw laser at 532 nm and a 30 mW cw laser at 850 nm, respectively. The 532 nm laser goes through an acousto-optic modulator (AOM), and we use the first-order diffracted beam for the experiment for reasons to be explained in the following. 

The two beams are then fiber-coupled into a VIS/NIR fiber combiner (Thorlabs NG71F1), the single output fiber of which has superimposed both colors on the same spatial mode. Thus both colors can be readily delivered to the experiment with common optics. However, the fiber introduces intensity and polarization fluctuations. A linear polarizer after the fiber output transforms the latter into the former. Since the experiment requires constant intensity, the intensity fluctuations need to be suppressed. To this end, we take advantage of the fact that the intensity of the diffracted beam at the 532 nm AOM is dependent on the rf power driving the AOM. By sampling the beam after the fiber combiner output with a photodiode, and using a stable voltage reference, we can create an error signal which is the input of a proportional/integral feedback. The output of the feedback is used to control a voltage-controlled attenuator, which controls the amplitude of an rf signal at 50 MHz. After proper amplification, the output of the voltage-controlled attenuator is fed into the rf input of the AOM, closing the feedback loop. 
\begin{figure*}
\begin{center}
\resizebox{0.95\textwidth}{!}{
 \includegraphics{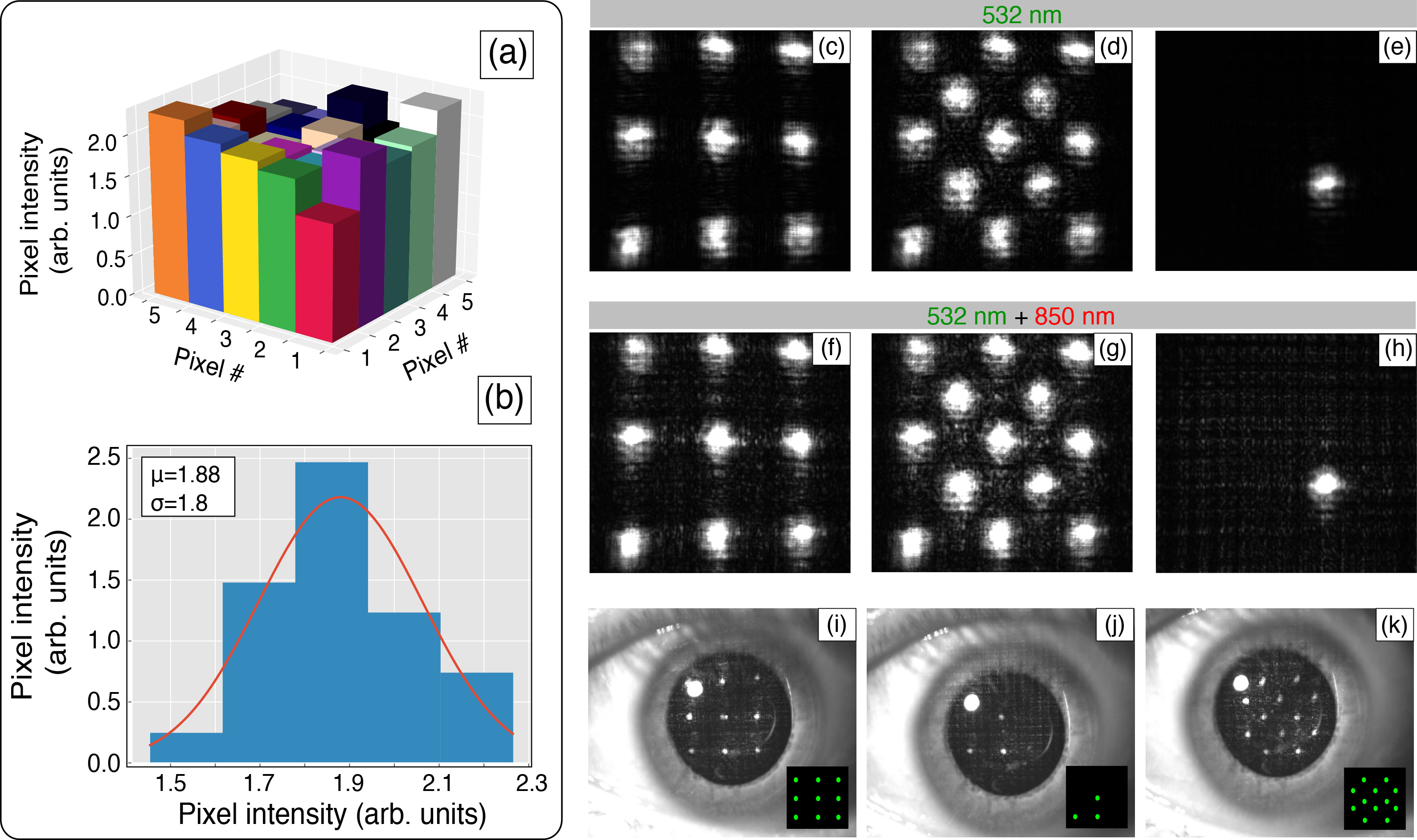}
}
\end{center}
\caption{(a) Uniformity of the intensity in the grid of $5\times 5$ pixels. (b) Distribution of intensities for all 25 pixels, showing a non-uniformity at the level of 10\%. (c,d) Result of superimposing green and infrared light with the fiber combiner. Shown here are two examples of arbitrary patterns of pixels illuminated by the 532 nm stimulus light, the grey scale image of which is superimposed on the image (red scale) obtained from the same patterns with the infrared pointing light at 850 nm. Both images were obtained by directly illuminating the CCD camera. In the images (e-g) we display the actual camera view of the corresponding author's left eye, for three examples of illumination patterns (for clarity the patterns are reproduced in the insets). The illuminated spots are visible through the infrared reflection on the glass plate preceding the eye (see Fig. \ref{setup}). The bright spot in the upper left part of the pupil is the reflection of the background infrared LED light used for imaging. }
\label{gir}
\end{figure*}

It is noted that the photodiode sampling the beam is illuminated by using a dichroic mirror, which reflects about 98\% of the 532 nm light exiting the fiber combiner into the photodiode and transmitts about 100\% of the infrared light. Thus the beam transmitted through the dichroic mirror has a power imbalance of the green/ir light at the level of 1/50. The role of the dichroic mirror will be further explained in Section 3.2. 

The beam transmitted through the dichroic mirror is then expanded to a diameter of about 3 cm, of which a nearly uniform-intensity central part of diameter 1 cm is directed to a graphic liquid crystal display (LCD). The purpose of the LCD is to create a desired pattern of illuminated pixels. The activated LCD dots are used to produce an optical loss in the incident beam, so that the pattern of activated/inactivated dots is transformed into a pattern of dark/illuminated pixels in the beam's cross section.
\subsection{Illumination pattern control}
In particular, the LCD has a grid of $192\times 64$ dots, each of dimension $0.458\times0.458~{\rm mm}$, and pitch $0.508\times0.508~{\rm mm}$. To avoid laser beam diffraction and abberation due to the 50 $\mu$m inactive groves between the dots we define a "pixel" for generating the $5\times 5$ grid of 25 illumination pixels to consist of a $2\times 2$ grid of LCD dots. That is, each of the 25 pixels has dimension $1.02\times 1.02~{\rm mm}$. With a higher optical quality LCD we can in principle significantly increase the number of pixels in the grid. 

With an activated LCD dot we obtain a relative optical loss of 0.35 for one-pass of the laser beam through the LCD compared to an inactivated LCD dot. Since for the biometrics protocol we wish to have photon numbers ranging up to 200 photons per illuminated pixel per pulse, a dark pixel should present a relative optical loss on the order of at least 1/200 in order for the photon number of the dark pixels to be negligible when the illuminated pixels have the maximum number of 200 photons. To obtain such a loss we use a 5-pass configuration produced by four right-angle prims, two on each side of the LCD as shown in Figs. \ref{setup}a,b. The combined optical loss is then $0.35^5=0.005\approx 1/200$. The LCD dots are programmed and the prisms are positioned in such a way that the same dot pattern appears at the position where the beam is incident on the LCD after being reflected by the prisms, so that indeed the optical loss is multiplicative with the number of passes.

An example of a LCD dot arrangement producing a single illuminated pixel is shown in Fig. \ref{setup}c, while an example with a pattern of $3\times 3$ next-to-nearest neighbor illuminated pixels is shown in Fig. \ref{setup}d. 
\begin{figure}
\begin{center}
\resizebox{0.46\textwidth}{!}{
 \includegraphics{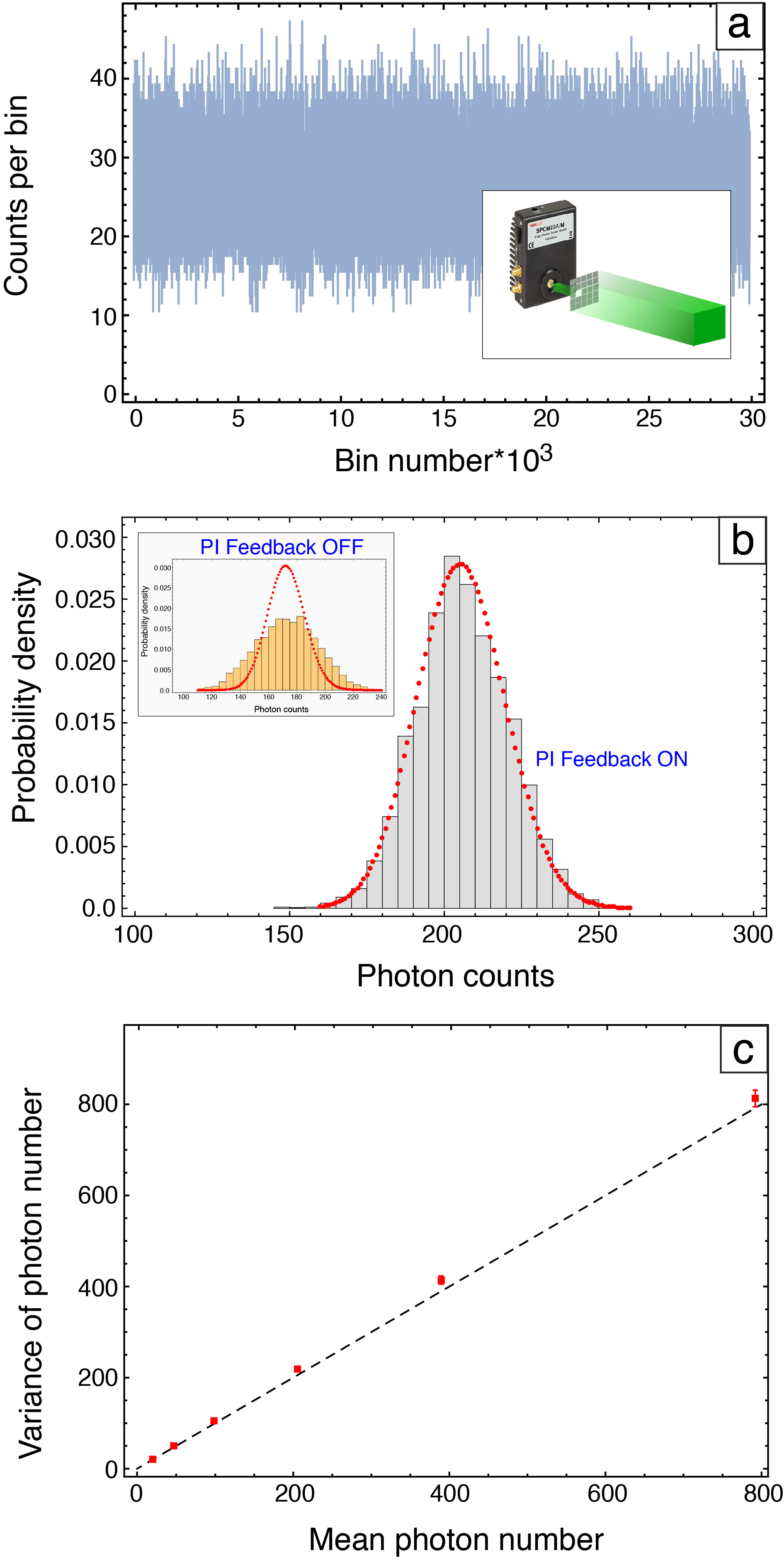}
}
\end{center}
\caption{(a) Count rate of the single-photon counter monitoring a single pixel of the $5\times 5$ illumination grid. As shown in the inset, all LCD dots but the ones corresponding to the monitored pixel are activated. Bin size is 12.5 ms. (b) Poisson distribution of the counts for one particular value of the mean number of counts. The counts correspond to 100 ms (8 bins) measurement intervals. The average value of the dark counts (about 1.4 counts per 100 ms) has been subtracted. Inset shows the broadening of the distribution when the 532 nm laser intensity exiting the fiber combiner is not actively stabilized. Red dots depict the Poisson distribution with the same mean as the counts histogram. (c) Variance $V_c$ versus mean $\overline{c}$ of the number of counts for several values of $\overline{c}$. Dashed line is $V_c=\overline{c}$. The error $\sigma_{V_c}$ of the variance is given by $\sigma_{V_c}^2\approx 2\overline{c}^2/N$, where $N=30000/8$ is the total bin number. The slight increase in the variance with the mean is due to the 2\% level of intensity noise after stabilization. It is irrelevant for our range of interest below 200 counts per 100 ms pulse. }
\label{PS}
\end{figure}
The dot pattern on the LCD is produced by a graphic user interface and then translated into a code driving the LCD's arduino controller. Thus, any kind of illumination pattern of the $5\times 5$ grid can be computer-generated, either by direct user input, or by some pre-programmed protocol. 

We have furthermore measured the uniformity of the intensity of the pixels. The result is shown in Fig. \ref{gir}a, showing the distribution of intensities in the $5\times 5$ grid. The relative non-uniformity is at the 10\% level, as shown in Fig. \ref{gir}b, which should be acceptable for the quantum biometrics protocol. 
\subsection{Superimposed infrared pointing light and green stimulus light}
A crucial aspect of our source, as compared to the state-of-the-art in pupillometry described in Section 2, is the fact that we use infrared light as pointing light conveying the exact spot of the stimulus on the pupil. This is achieved by the fiber combiner (Fig. \ref{setup}), the output of which contains both the green stimulus light and the infrared pointing light (not perceived by the human visual system) in the same spatial mode of the laser beam exiting the fiber combiner. 

The result of this superposition is shown in Figs. \ref{gir}c,d, where we show two examples of various illumination patterns. Images were acquired by the laser beam exiting the LCD directly illuminating a CCD camera, first with 532 nm light only, and then with 850 nm light only. The obtained images were then superimposed with different color coding. The perfect spatial superposition is readily observed by inspecting Figs. \ref{gir}c,d.

The final optical element before the eye is a glass plate. Its purpose is to reflect the infrared light of the beam through the beam splitter back to the CCD camera (see Fig. \ref{setup}), providing information on the exact location of the illuminated spots on the pupil. This is shown in Figs. \ref{gir}e-g for three different examples of illumination patterns. We can now explain the role of the dichroic mirror mentioned previously. Applying the light stimulus in the biometric identification protocol requires pulses of light containing 20-200 photons at 532 nm. If the infrared and the green light are of similar power, reducing the power of the 532 nm light exiting the fiber combiner to a level such that 20-200 illuminate the eye will do so also for the infrared light, hence its reflection will not be visible. On the other hand, reducing the 532 nm light power before entering the fiber combiner will diminish the signal of the photodiode and deteriorate the feedback loop. The dichroic mirror ameliorates both of these problems.
\subsection{Photon number control and photon statistics}
As already mentioned, in order to scan the probability of seeing curve, we need to be able to change the incident photon number per pulse from about 20 to 200 photons. To do so we use the feedback circuit that stabilizes the 532 nm laser intensity at the output of the fiber combiner. The stabilization rests on using the PI feedback circuit to lock the photodiode signal $V_{\rm pd}$ to a stable voltage reference (see Fig. \ref{setup}). This reference can be provided by the analog output, $V_{\rm ref}$, of a DAQ card, and can readily change by a factor of 10, e.g. from 10V to 1V. We can thus set the 532 nm light intensity so that 200 photons per pulse per illuminated pixel correspond to $V_{\rm pd}\approx 10~{\rm V}$. Then by changing $V_{\rm ref}$ from 10V to 1V we can obtain pulses containing 200 down to 20 photons on average. 

To measure the statistical distribution of the photon number in an illuminated pixel of 532 nm light, we focus the light of such a pixel into a single photon counter (Thorlabs SPCM20A). An example of the count rate is shown in Fig. \ref{PS}a. We then measure the distribution of the counts in 100 ms time intervals for various values of the mean photon count. We find a Poisson distribution for all values of the mean that are of interest. An example is shown in Fig. \ref{PS}b. We note that without the active stabilization of the 532 nm light exiting the fiber combiner, the photon count distribution is, as expected, much wider than a Poisson distribution, as shown in the inset of Fig. \ref{PS}b. Finally, in Fig. \ref{PS}c we plot the variance of the measured distributions as a function of the mean for a mean count up to 800, displaying the Poisson distribution for the region of interest. It is noted that these measurements represent the actual counts of the single photon counter and have not been calibrated (using the various losses and efficiencies) to reflect absolute photon counts. This is an exercise to be undertaken before the actual illumination of the human subjects in a biometric protocol.
\section{Spatially-selective pupillometry}
We will here provide proof-of-concept pupillometry data demonstrating the potential of the light stimulus source to extract new physiological information due to its spatial selectivity. We will not venture into any physiologial interpretation of the data, but just qualitatively demonstrate the wealth of information that can be in principle extracted from the measurements. 
\subsection{Pupil diameter measurement}
\begin{figure*}
\begin{center}
\resizebox{0.95\textwidth}{!}{
 \includegraphics{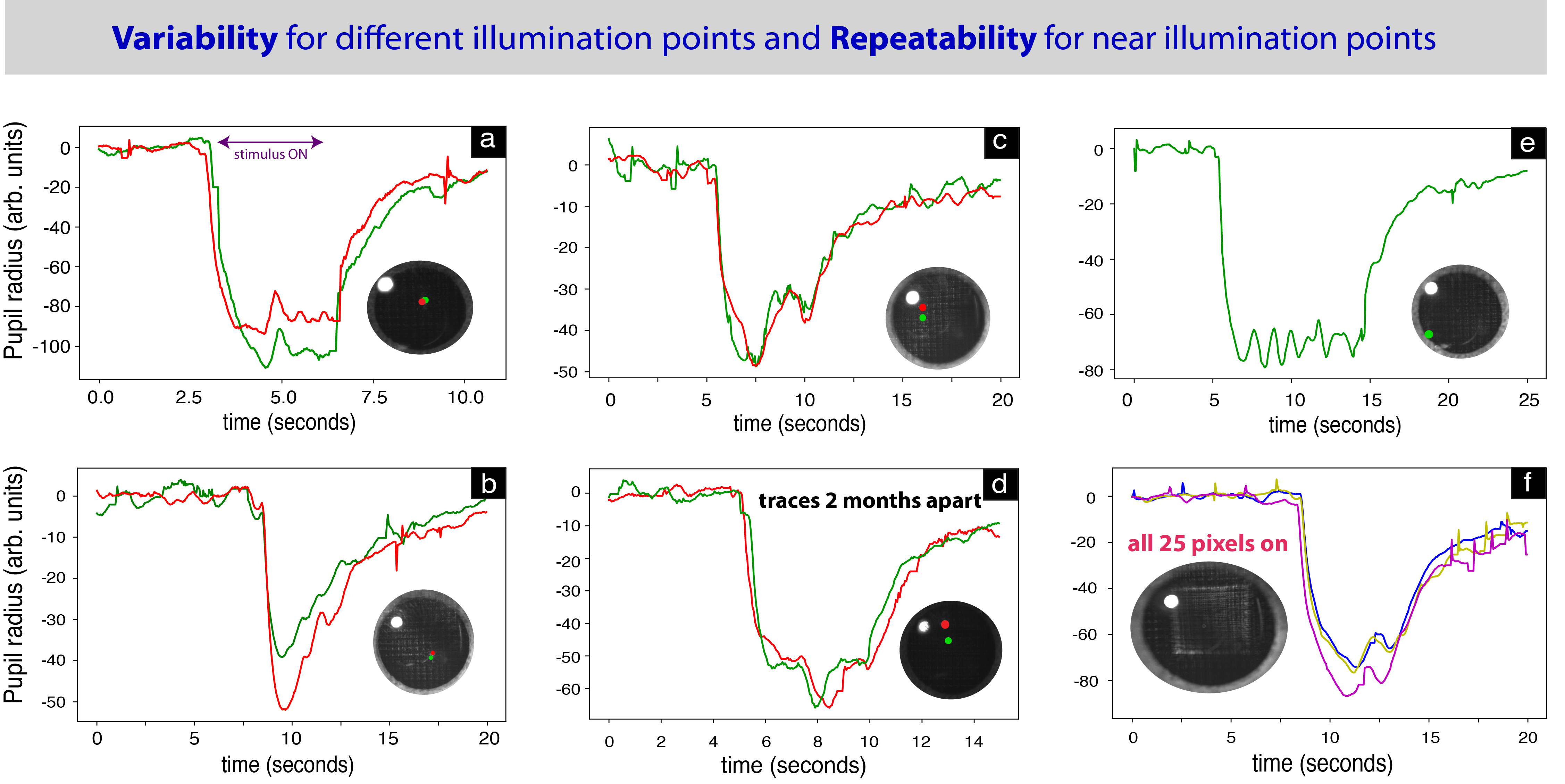}
}
\end{center}
\caption{Spatially selective pupillometry measurements. (a-d) When same or closely spaced spots on the pupil are illuminated, the pupil diameter exhibits very similar response. For reasons outlined in the text, the repeatable responses presented here represent 20\%-30\% of all measured responses. In the data shown in Fig. 10 this rate is improved to 50\% with an alternate "smart" data taking methodology including a neural network. (d) These two traces were taken 2 months apart, pointing to objective features of the pupil light reflex measurements. (e) In this case the illumination spot was by chance right on the pupil's boundary, so that the pupil's constriction decreases the light coming in, prompting the brain to re-dilate the pupil, and so forth, leading to the observed oscillations. (f) In this case, which can serve as "control", all 25 pixels are illuminated, producing a square shaped beam seen by the infrared reflection. Thus, illuminating the whole pupil can provide a single response, whereas we can in principle extract additional information from 25 different responses when illuminating 25 different pixels. It is clear that there is rich information to be extracted from this spatially selective single-pixel illumination (e.g. shape of responses, timing of their features), as opposed to illuminating the whole pupil indiscriminately. All measurements were performed on the corresponding author's left myopic eye, which leads to the perception of different pixels, since a normal eye would focus all parallel beams to a single spot on the retina.}
\label{RV}
\end{figure*}

For measuring the pupil response to illumination, as presented in the following, we use illumination of a single pixel chosen out of the 25 pixels we have available. In contrast with the biometrics protocol, in the pupillometry measurements the photon number per unit time is much larger. In particular, for each pixel we have about 40 M photons/s/pixel incident on the eye, for a duration described with the horizontal bar in Fig. \ref{RV}. 
\begin{figure}
\begin{center}
\resizebox{0.45\textwidth}{!}{
 \includegraphics{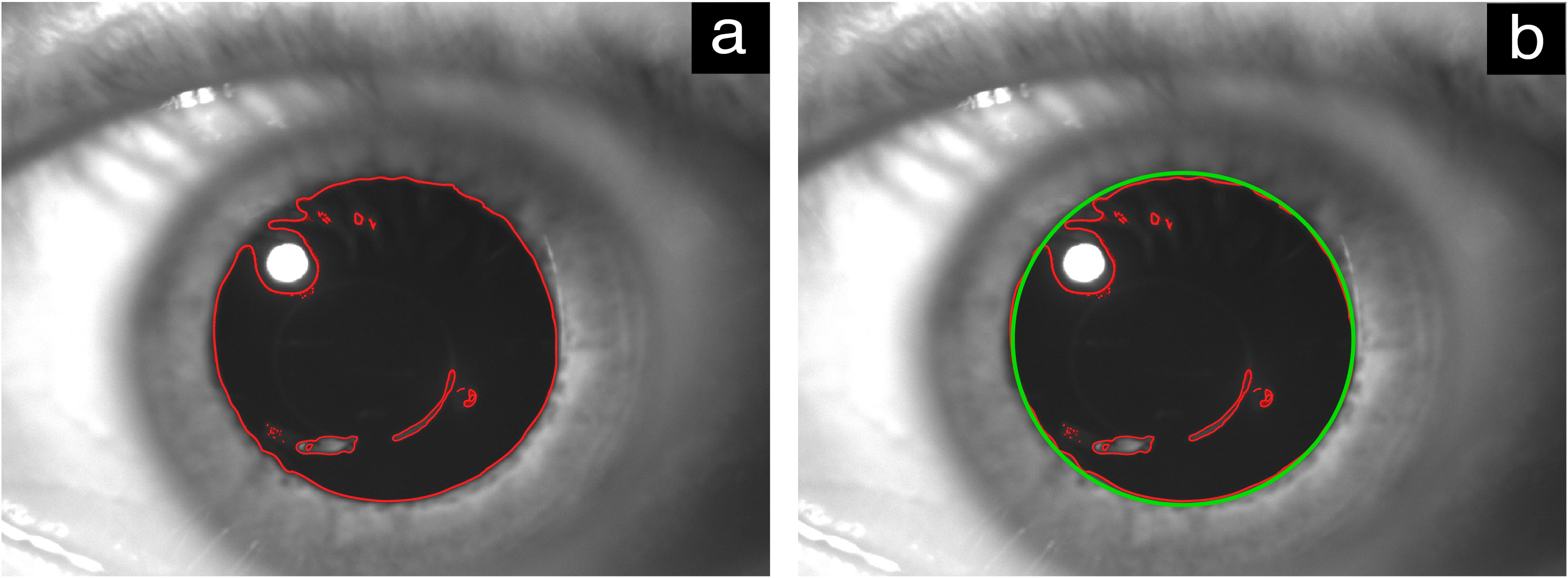}
}
\end{center}
\caption{(a) Contours defined by the image's intensity gradient as described in the text. (b) Maximum surface contour defines the pupil. The green circle is defined by the two furthest points of the maximum surface contour.}
\label{pd}
\end{figure}

To extract the pupil diameter as a function of time, we analyze the image frames captured by the CCD camera, the frame rate of which was 20 fps. In some cases we acquired 350-frame data, and in some other cases 500-frame data. Each image frame was slightly blurred to reduce noise, and a greyscale intensity gradient was defined based on the difference between the black pupil and the light-grey iris. The definition of this gradient depends on the ambient light conditions, i.e. although we have a fixed source of infrared LED light illuminating the whole eye, small changes in the position of some optics or small changes in the head's position affect this background light conditions. For each set of frames this gradient definition was optimized by visually inspecting the resulting circular fits to the pupil. The latter were obtained after forming the contours defined by the intensity gradient, the maximum surface contour corresponding to the pupil. The pupil diameter is the one resulting from the circle defined by the two furthest points on the contour. The aforementioned contours and the resulting circular fit to the pupil are shown in Fig. \ref{pd}a and Fig. \ref{pd}b, respectively.
\subsection{Variability and repeatability and of the pupil responses under spatially selective illumination}
A significant variability of the pupil response for illumination of different spots on the pupil is demonstrated Figs. \ref{RV}a-d. From these responses it is evident that there is very rich and qualitatively different information to be extracted from the responses following illumination of different spots. The infromation is e.g. about the specific features of the response, their timing, the velocity (time derivative) of the pupil diameter variation etc. One distinct example demonstrating the wealth of infromation that can be extracted is Fig. \ref{RV}e, which exhibits clear oscillations during the illumination period. This is because the illumination point happened to be right on the boundary of the pupil. When illuminated, the pupil contracts and the received light almost disappears, prompting the brain to re-dilate the pupil, and so forth. It is e.g. expected that the oscillation period and amplitude can be connected to fundamental properties of the relevant neural circuits. 

For the sake of having a "control" measurement, we turned on all 25 pixels forming an illumination square. Its infrared reflection, along with the corresponding repeatable response is shown in Fig. \ref{RV}f. Whatever the response, and whatever the mechanism producing such a response from a broad illumination, it is obvious that this represents just one type of response carrying limited information compared to the full array of responses following spatially-selective illumination of 25 pixels in principle.

As mentioned before, it is not the purpose of this proof-of-concept work to venture into physiological interpretations of these data. However, it is not a a far stretch to assume that the fundamental source of the observed variability in Figs. \ref{RV}a-d  is that illumination of different spots on the pupil translates into stimulation of different spots on the retina, hence excitation of different neural circuits regulating the pupil's responses. Again, this is due to the fact that the subject's eye is myopic, and hence the parallel light beams illuminating the pupil are focused on different spots on the retina. For emmetropic eyes one can in principle use a diverging lens before the eye to produce a similar effect. 

\begin{figure}
\begin{center}
\resizebox{0.4\textwidth}{!}{
 \includegraphics{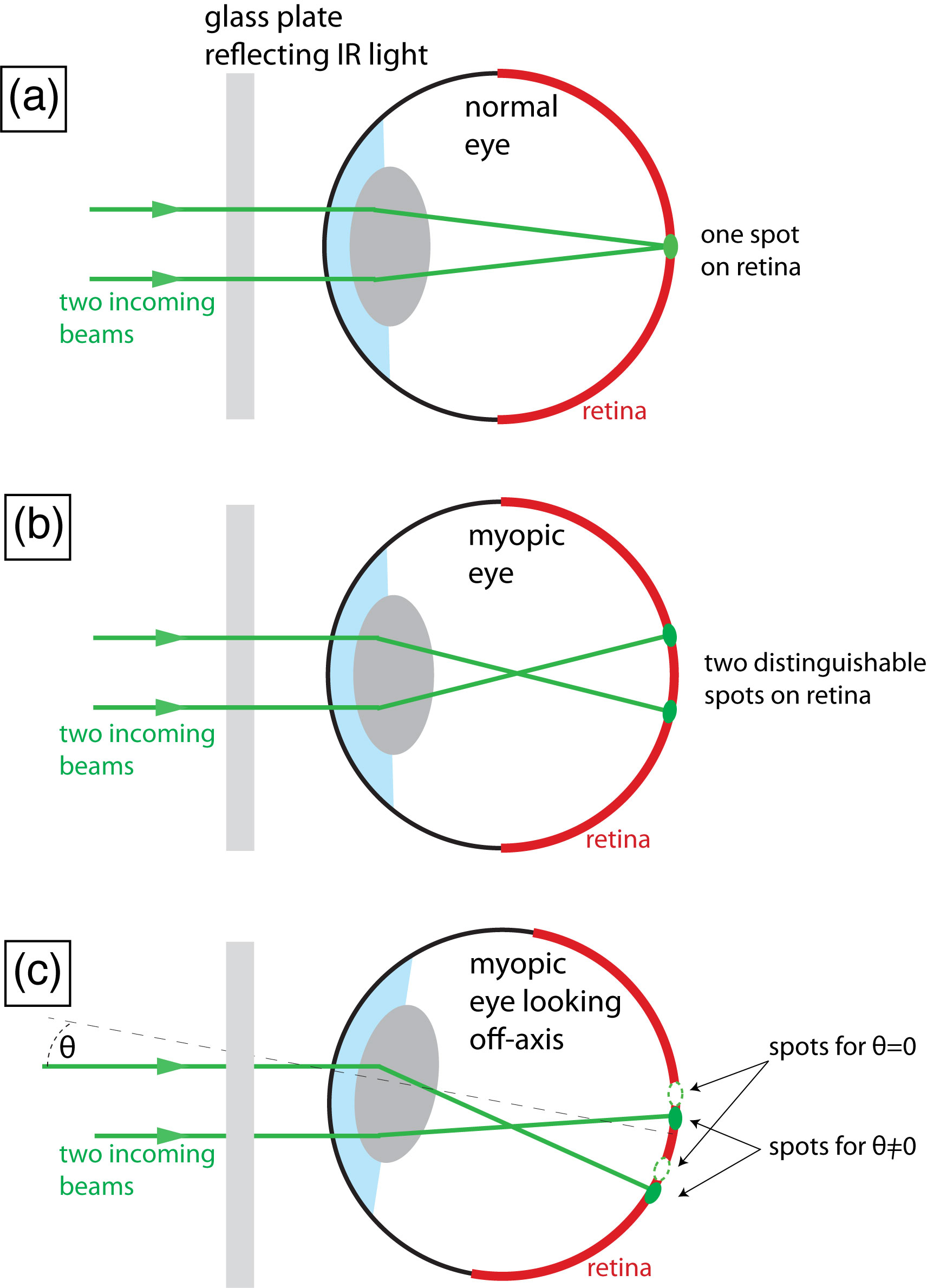}
}
\end{center}
\caption{Geometry of eye's illumination. (a) For a normal eye, two parallel beams incident on a different spot on the pupil are focused on the same spot on the retina. (b) For a myopic eye, as the subject's eyes used for the measurements in this work, two parallel beams incident on a different spot on the pupil will be perceived as two distinct spots, because the focus is before the retina, hence they will indeed illuminate two different spots on the retina. (c) Example showing that a small rotation of the head/eye with respect to the laser beam's axis will result in different spots being illuminated in the retina, whereas the same spot will be seen to be illuminated on the glass plate preceeding the eye. Approximately, just $1^{\circ}$ of rotation results in a change in focal spot position by $\delta x\approx \theta d\approx 0.5~{\rm mm}$, where $d=24~{\rm mm }$ is the typical eye diameter and $\theta=\pi/180$. Already for such small $\theta$, it follows that $\delta x$ is quite larger than the spot diameter stimulating the retina as the following rough calculation can show. For a  beam waist radius at the focus of $w_0\approx 10~{\rm \mu m}$, and a refractive index of 1.4, the Rayleigh range $z_R\approx 0.8~{\rm mm}$. Taking the focus of the myopic eye to be roughly at the center of the eye, the distance from the focus to the retina will be $z\approx 12~{\rm mm}$, hence the focal spot on the retina will be $w\approx (z/z_R)w_0\approx 150~{\rm \mu m}$.}
\label{geometry}
\end{figure}

Regarding the repeatability of the measurements in Figs. \ref{RV}a-d, it is again reasonable to assume that it stems from different measurements stimulating the same (or nearby) spots on the retina. However, we here need to clarify several subtle points.\newline
{\bf (i)} Our head and chin rest does not allow precise and repeatable positioning of the head/eyes with respect to the fixed illumination beam. This can lead to the following perplexing situation. As explained in Fig. \ref{geometry}, it does not necessarily mean that whenever we observe nearby spots {\it on the pupil} (through the ir reflection), we will be stimulating the same spot {\it on the retina}. Converesly, it does not necessarily mean that whenever we observe that we illuminate two different spots on the pupil, we will be illuminating two different spots on the retina. This is the case because, as shown in Fig. \ref{geometry}, a slight relative rotation or translation of the head/eye with respect to the laser beam direction within a set of, say two different measurements, might lead to two different spots (or for that matter a single spot) on the retina, even if it appears we are illuminating the same spot (or for that matter two different spots) on the pupil.\newline
{\bf (ii)} The result of this is that the variable responses in Figs. \ref{RV}a-d are not in a 1-1 correspondence with the pupil illumination point as observed from the infrared reflection. Moreover, the repeatable responses shown in each of the Figs. \ref{RV}a-d are only 20\%-30\% of the total number of responses acquired for each nominally selected illumination pixel. \newline
{\bf (iii)} Due to the lack of an obvious  1-1 correspondence  mentioned in (ii), one could claim that we subselect data that randomly manifest some repeatabilty. One first counterargument is Fig. \ref{RV}d, where the two almost identical responses where acquired at two different times 2 months apart! Moreover, when talking about repeatability we should stress that we are not comparing just numbers, but whole time traces having several distinct features. Hence it is not conceivable how such repeatability could be produced randomly without the underlying foundamental cause presented before. It can thus be stated that the fact that we do observe repeatable responses even at the level of 20\%-30\%, coupled with the argument in (i) is enough to make the case that indeed the variability (repeatability) of the responses is due to the illumination of different (nearby) spots on the retina.

To strengthen this argument, we have developed a new method of "smart" data taking, which uses a deep neural net to perform a preliminary form of eye tracking and "lock" the illumination point on the pupil between the consecutive data sets. Thus we managed to boost the repeatablity of the measured responses to the level of 50\%.
\subsection{Eye tracking}
\begin{figure}
\begin{center}
\resizebox{0.4\textwidth}{!}{
 \includegraphics{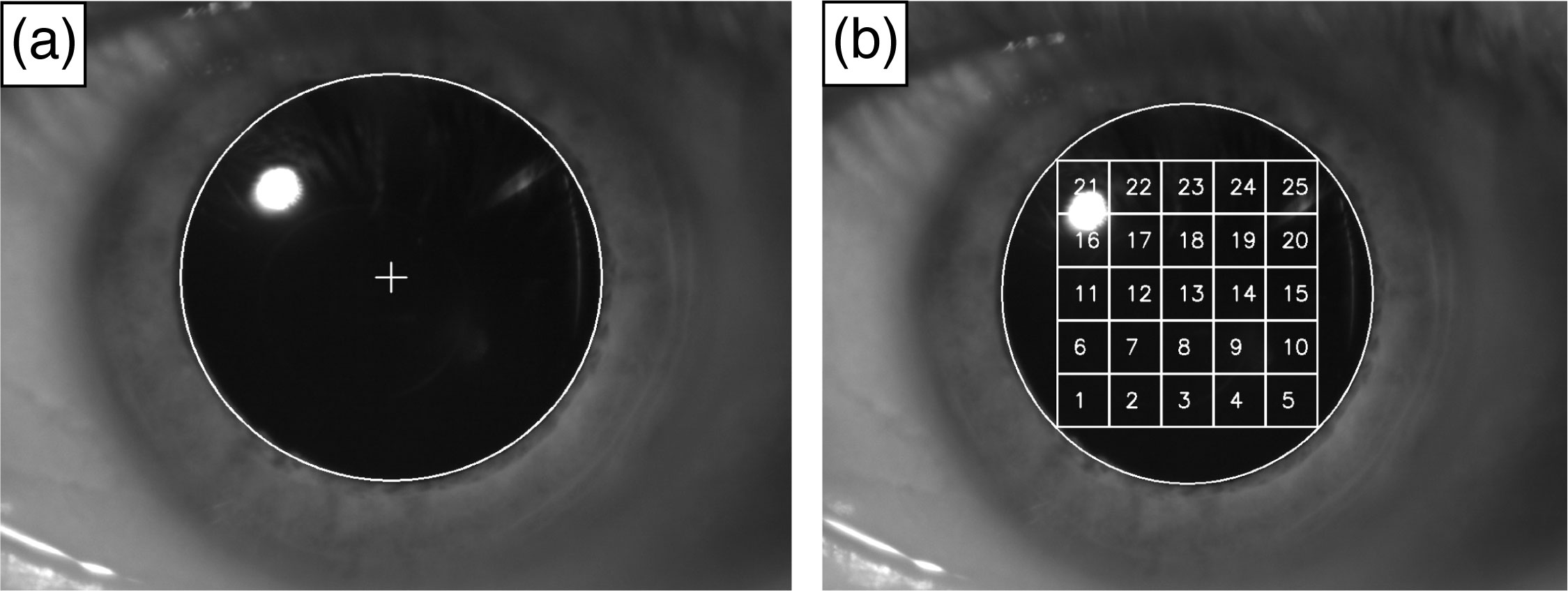}
}
\end{center}
\caption{(a) Determination of the center and diameter of the pupil using the first 20 frames in the dark. (b) Definition of a grid of 25 boxes to chose from for the nominal illumination point.}
\label{tracking}
\end{figure}

The way this works is the following. Using the infrared reflection from the glass plate preceeding the eye, we register the coordinates of the 25 illumination pixels produced by the LCD. As soon as the subject is in position for the first measurement, we register without any stimulus light 20 frames of the pupil, in order to determine its center and its diameter (Fig. \ref{tracking}a). We then define a grid of 25 boxes on the pupil as shown in Fig. \ref{tracking}b, and decide which box will be the nominal illumination box. The computer software then finds which LCD pixel is closest to the desired illumination box, and illuminates that particular LCD pixel. In subsequent measurements, for which the head/eye might have moved with respect to the laser beam the above procedure is repeated to find the LCD pixel closest to the nominal illumination box. For those measurements during which the subject's eye has remained largely in a fixed position relative to the laser beam, we find that the applied correction is mostly 1 LCD pixel. For the cases where the subject shortly leaves the chin rest and returns, the applied correction might be several LCD pixels. 

Moreover, we have trained (see below) a neural net to classify the images into two cases, eiher "acceptable", for which the pupil is seen unobstracted (e.g. from the eye lid), or "unacceptable", for which e.g. the eye lid is momentarily closed. This not only helps the extraction of the pupil diameter from all the frames of a measured response, but it also provides for a well-defined initial position and diameter using the first 20 frames in the dark mentioned before. 

In particular, the deep neural net was developed using the Keras Deep Learning Library. It is comprised of 7 layers of neurons (the input layer, 5 hidden layers and 1 output layer), the first 6 of which with 50 neurons each. The last layer, since this is a classification network, has only 1 neuron being activated by the Sigmoid function, classifying the image as "acceptable" or "unacceptable". The input layer consists of 50 neurons being activate by the rectified linear unit function. The input of the network is a scaled down image, from the original resolution of $1280\times 1024$ pixels to the smaller resolution of $100\times 80$ pixels. The neuros in the hidden layers are activated using the tanh function. The training set for the network was created from a series of about 3000 images of the subject's left eye, which were classified by the authors as "acceptable" or "unacceptable". After training (using 80\% of the total dataset) the network reached a success rate of 99.6\% on classifying the images of the test set (20\% of the total dataset). 
\begin{figure}
\begin{center}
\resizebox{0.38\textwidth}{!}{
 \includegraphics{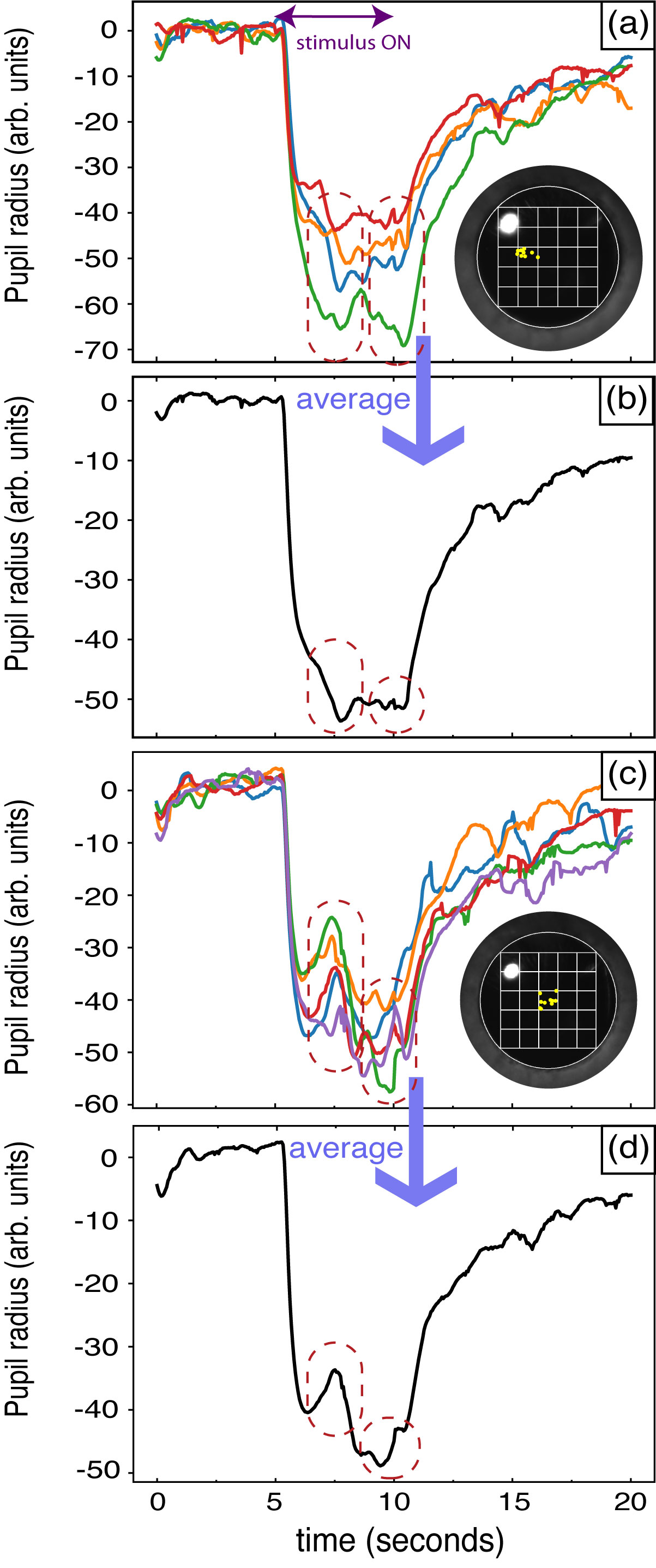}
}
\end{center}
\caption{(a) and (c) Pupil diameter responses for nominal illumination of box 12 and 13, respectively. Yellow circles are the centers of the illumination points inferred from the IR reflection. (b) and (d) are averages of (a) and (c), respectively. Dotted boxes highlight the distinct and conserved features of the responses, which are clearly different between cases a,b and c,d. Compared to tha data of Fig. 7, where repeatability was at the level of 20\%-30\%, here the repeatable responses were 50\% of all measured responses.}
\label{repeat2}
\end{figure}
\subsection{Boosted Repeatability}
\begin{figure}
\begin{center}
\resizebox{0.4\textwidth}{!}{
 \includegraphics{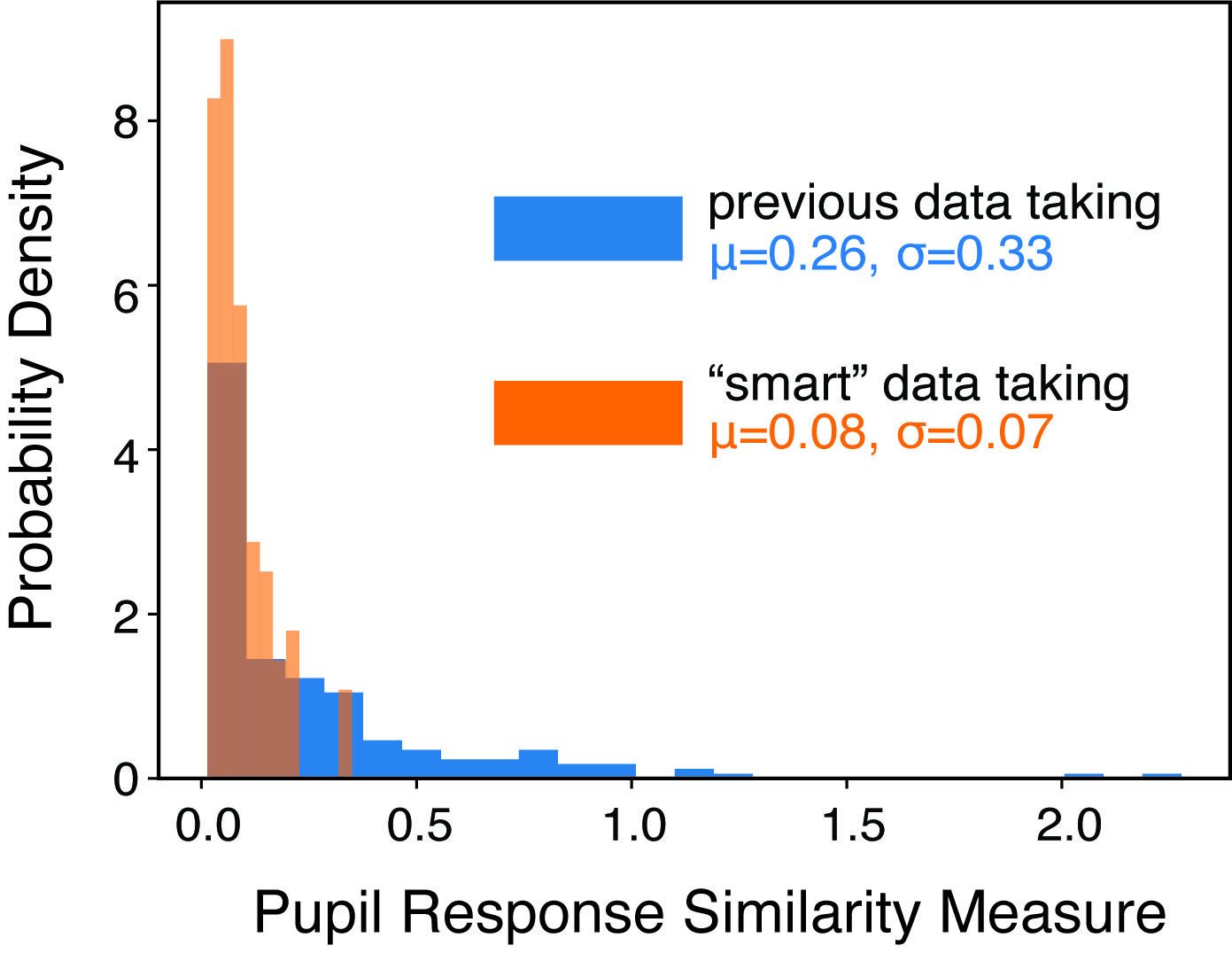}
}
\end{center}
\caption{Distribution of $\chi$ measuring the similarity of the pupil responses, showing the increased similarity obtained (suppression of $\chi$) with the "smart" measurement scheme.}
\label{corr}
\end{figure}

The results of this data taking method are shown in Fig. \ref{repeat2} and Fig. \ref{corr}. First, in Figs. \ref{repeat2}a,b we present an example of illuminating box 12 and in Figs. \ref{repeat2}c,d an example of illuminating box 13 (see Fig. \ref{tracking}b for the definition of the boxes). In the former case we measured 9 responses in total, and in the latter case 10 responses. Plotted in Figs. \ref{repeat2}a,c are the responses exhibiting repeatable features, so it is evident that the repeatablity has increased to about 50\%. In the insets of Figs. \ref{repeat2}a,c we can observe the centers of the illumination pixels in each measurement resulting from this "active" eye-tracking methodology. In Fig. \ref{repeat2}b and Fig. \ref{repeat2}d we depict the average of all traces in Fig. \ref{repeat2}a and Fig. \ref{repeat2}c, respectively, highlighting (dotted boxes) the distinct and conserved features in each case. Again, it is clear that moving the illumination point by one box results in noticeably different and repeatable features in the pupil diameter response. 

Finally, instead of qualitatively observing the repeatable responses, we calculated a "similarity" measure of all pairs of responses in each set, defined by $\chi=\int dt[a(t)-b(t)]^2/\sqrt{\int dt a(t)^2\int dt b(t)^2}$, where $a(t)$ and $b(t)$ is the pupil radius as a function of time for responses a and b, respectively. Obviously, the more similar are the responses $a(t)$ and $b(t)$, the smaller is $\chi$. For this calculation we used only the part of the responses during the illumination period of 5 seconds, since it is within this period that the variability of the responses for various illumination points is most pronounced and most interesting. In Fig. \ref{corr} we show the distribution of $\chi$ for the previous and current data taking mode, clealy obtaining a factor of 4 reduction in the mean and standard deviation of $\chi$, which is consistent with the observed boost in repeatablity. 

As a final remark, with this methodology we solve the problem of parallel translations of the head/eye on the plane transverse to the laser beam. We have not solved the problem of rotations with respect to the beam axis, e.g. as shown in Fig. \ref{geometry}. 
Nevertheless, we have presented several arguments supporting our basic claim that the more systematically we can stimulate the same spot on the retina, the more repeatable responses will be obtained.

\section{Conclusions and outlook}
Using modern tools of photonics technology, we have presented a light stimulus source capable of providing a coherent light beam in the optical spectrum, the cross section of which consists of discrete pixels, the illumination of an arbitrary pattern of which is computer controlled. Moreover, the photon number per unit time per pixel is precisely known, as it is distributed according to Poisson statistics describing coherent light.  The mean photon number per unit time per pixel can be chosen at will within a range of at least an order of magnitude, also by computer control. Additionally, infrared light is exactly superimposed in the same beam as the stimulus light in the optical spectrum, so that it can provide precise information on the point of illumination through reflection. 

The light stimulus source reported herein can be used to realize protocols for biometric identification using the visual system's ability to perform photon counting. Furthermore, it can be used to study the pupil reflex dynamics with spatial selectivity of illumination, and thus selectively probe neural circuits, which so far was not possible, as current pupillometry indiscriminately illuminates the whole pupil with classical light. We have shown that with the spatially selective stimulation we can in principle obtain so far inaccessible information, which could help diagnose several medical conditions relevant to vision as well as the central nervous system. 

Finally, we project that a more advanced version of this light stimulus source can be developed by using quantum light instead of coherent light, possibly leading to a quantum-enhanced understanding of the human visual system.

This work has been co-financed by the European Union and Greek national funds through the Operational Program Competitiveness, Entrepreneurship and Innovation, under the call RESEARCH - CREATE - INNOVATE, with project title "Photonic analysis of the retina's biometric photoabsorption" (project T1EDK - 04921).

\end{document}